# Asymmetric Band Diagrams in Photonic Crystals with a Spontaneous Nonreciprocal Response

Filipa R. Prudêncio [(1), &], Sérgio A. Matos[(2)] and Carlos R. Paiva[(1)]

[1]*Department of Electrical and Computer Engineering – Instituto de Telecomunicações,*
*Instituto Superior Técnico – University of Lisbon*
*Avenida Rovisco Pais, 1, 1049-001 Lisboa, Portugal*

[2]*Department of Information Science and Technology – Instituto de Telecomunicações*
*Instituto Universitário de Lisboa – ISCTE*
*Avenida das Forças Armadas, 1649-026 Lisboa, Portugal*

## Abstract

We study the propagation of electromagnetic waves in layered photonic crystals formed by materials with a spontaneous nonreciprocal response, such as Tellegen (axion) media or topological insulators. Surprisingly, it is proven that stratified Tellegen photonic crystals that break simultaneously the space inversion and time reversal symmetries have always symmetric dispersion diagrams. Interestingly, we show that by combining chiral and nonreciprocal materials the photonic band diagrams can exhibit a spectral asymmetry such that $\omega(\mathbf{k}) \neq \omega(-\mathbf{k})$. Furthermore, it is demonstrated that in some conditions two juxtaposed Tellegen medium layers have an electromagnetic response analogous to that of a biased ferrite slab.

---

[&] Corresponding author: filipa.prudencio@lx.it.pt

# I. Introduction

Photonic crystals are inspired by the geometry of crystalline natural materials, such as semiconductor crystals [1], [2]. Periodic layered dielectric structures are of particular importance in the realization of filters, resonant cavities, and light sources [3], [4]. The propagation of light in crystals with nonreciprocal media has also received considerable attention in the literature [5]-[7]. In particular, recent works have demonstrated how by exploring interference phenomena in different magneto-optical photonic crystalline structures it is possible to boost the nonreciprocal response and achieve optical isolation, optical switching, one-way extraordinary optical transmission, and a giant Faraday rotation [5]-[17]. These efforts are partly motivated by the contemporary interest in on-chip miniaturization of all-optical circuits, which requires on-chip optical signal isolation.

Notably, having a nonreciprocal response and asymmetric power flows is far from trivial in photonics. Indeed, the Lorentz reciprocity theorem establishes that a two-port network formed by conventional dielectrics and metals is intrinsically bi-directional, such that the roles of the source and load can be interchanged without affecting the amount of power transmitted through the system. This contrasts with electronics wherein the rectification functionality is provided by diodes and transistors, which effectively behave as unidirectional couplers for electrons. A nonreciprocal response can be obtained with an external bias magnetic field (e.g. using ferrimagnetic materials such as ferrites at microwaves [5] or with bismuth iron garnet at optics [8]) with the temporal refractive-index modulation [9], or by using optomechanical effects (e.g. with moving media [10]). Even though the first solution is quite established, the requirement of an external biasing is inconvenient because it limits the range of applications and because the associated biasing circuit may be bulky. In general, devices made of nonreciprocal media can allow for highly asymmetric transmissions [5]-[7]. A related effect can also occur in more restricted conditions in chiral structures [18]-[19]. Moreover, photonic crystals made of chiral media may be useful for polarization state conversion [18], [19].

Tellegen media are a subclass of bi-isotropic media with magnetoelectric coupling [20], [21]. The key characteristic of Tellegen media is the combination of an isotropic response with a *spontaneous* nonreciprocal magnetoelectric effect which [20], *does not* require any external biasing. This idea was put forward by Tellegen in 1948, in connection with his proposal of a new nonreciprocal circuit element [22]. He suggested that a mixture of randomly distributed particles with glued permanent electric and magnetic dipole moments may have a magnetoelectric isotropic response. Some time ago an artificial Tellegen medium was implemented using essentially this recipe [23]. Notably, the physics of the Tellegen medium is a particular form of axion electrodynamics [24]. Axions were originally introduced by F. Wilczek as an attempt to explain the missing dark matter of the universe [25]. As noted by several authors, the magnetoelectric coupling due to a axion field with a time independent axion-coupling term is equivalent to the Tellegen constitutive relations [24], [26]-[28]. For completeness, we provide an explicit proof of this result in the Appendix A.

There are naturally available media with a spontaneous nonreciprocal magnetoelectric response, being the most prominent example chromium oxide $Cr_2O_3$ [29]. It was experimentally verified that the spontaneous magnetoelectric effect in $Cr_2O_3$ creates a polarization rotation in the reflected wave, which is the fingerprint of a nonreciprocal magnetoelectric coupling [30]. In the last decade there has been a resurgence in the interest in materials with a strong magnetoelectric response [31]-[34]. In particular, it has been suggested that the new class of crystalline solids known as electronic topological insulators [32], [33], may be characterized by axion-type electrodynamics, which as previously discussed is equivalent to the Tellegen medium response [28]. In most topological insulators an observation of the magnetoelectric effect may require adding a time-reversal breaking perturbation, e.g. a permanent magnet [32], i.e. a biasing element. Interestingly, it has been suggested that antiferromagnetic topological insulators can enable a spontaneous magnetoelectric effect, similar to $Cr_2O_3$ [34]. Moreover, in [31] it was theoretically demonstrated that there are several crystal structures and corresponding arrangements of magnetic moments compatible with a purely isotropic linear spontaneous nonreciprocal magnetoelectric coupling. Therefore, it is relevant to explore possible applications of materials with an isotropic spontaneous nonreciprocal response in the context of electromagnetic wave propagation. It is important to make clear that the topological magnetoelectric effect discussed above is totally unrelated to "photonic topological insulators", which have received significant attention in the recent literature [35]-[36]. The topological magnetoelectric effect is a consequence of the nontrivial topological nature of the *electronic* band structure of a bulk solid-state material [33]. Moreover, in this article we do not attempt to characterize the topology of the band structure of Tellegen photonic crystals.

Motivated by the above discussed developments in material science, here we investigate the consequences of a spontaneous nonreciprocal response in stratified periodic structures. The isotropy of a Tellegen medium implies that the dispersion of the photonic states is doubly degenerate and symmetric, so that $\omega(\mathbf{k}) = \omega(-\mathbf{k})$. Thus, in a certain sense the nonreciprocal response is hidden. Only in composite structures formed by different materials (e.g. a photonic crystal) it is possible to unveil the nonreciprocal character of the Tellegen media. In this work, we want to explore under which conditions it is possible to have asymmetric band diagrams in photonic crystals with Tellegen media and which physical symmetries of the photonic crystal need to be broken to achieve this.

Surprisingly, it is found that the band structure of layered Tellegen photonic crystals is always symmetric for propagation along the stratification direction, notwithstanding the space inversion and time reversal symmetries are broken. We explain this result by identifying another symmetry transformation that protects the spectral symmetry of the band diagrams. Importantly, it is shown that by putting together chiral media and Tellegen media, it is possible to obtain asymmetric band diagrams, $\omega(\mathbf{k}) \neq \omega(-\mathbf{k})$, and highly asymmetric group velocities. Furthermore, we prove that in some conditions the role of two Tellegen layers in the photonic crystal is essentially equivalent to a biased ferrite layer. We would like to highlight that other authors have shown that a proper arrangement of ferrites and anisotropic dielectrics may yield asymmetric band diagrams [5]-[6]. However, here we demonstrate for the first time that the same result can be obtained without a biasing static field based on media with a spontaneous nonreciprocal response.

This article is organized as follows. In Sect. II we start by characterizing the nonreciprocal effects in the scattering by Tellegen stratified structures using the scattering matrix formalism [37], [38]. In Sect. III we present a similar study for the case of stratified ferrite structures. In Sect. IV we apply the developed formalism to determine the band structures of Tellegen and ferrite photonic crystals. Finally, in Sect. V the conclusions are drawn.

## II. Stratified Structures with Bi-isotropic Media

As a starting point, we characterize the scattering of waves by stratified structures formed by bi-isotropic media. The geometry of the structure is represented in the Fig. 1a wherein each slab is either a Tellegen material, or a chiral material or a conventional dielectric. The stratification direction is assumed to be the *z* direction and the slabs have thicknesses $d_1, d_2, ..., d_N$. The longitudinal propagation constants (along the *z*-direction) are denoted by $\beta_1, \beta_2, ..., \beta_N$. The input and output regions are assumed to be the vacuum or air.

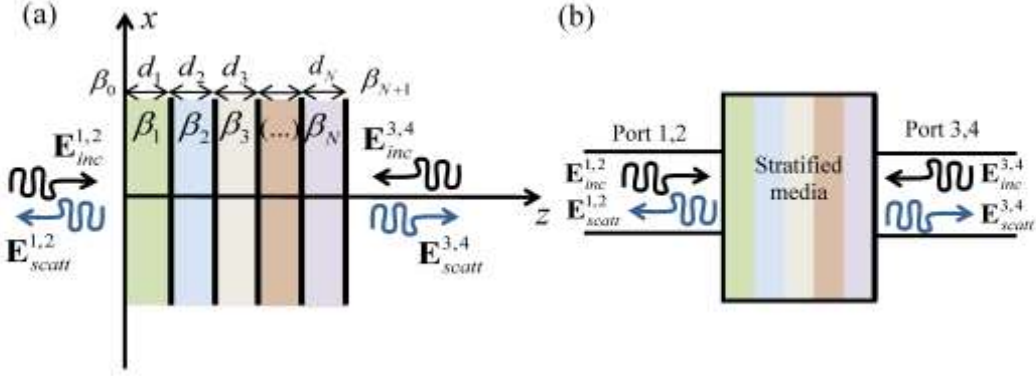

**Fig. 1** (Color online) The system represented in panel (a) is equivalent to the four-port network shown in panel (b). Even though there are only two physical channels, because of the polarization degrees of freedom the equivalent network is formed by four ports.

The electromagnetic response of bi-isotropic media is determined by the well-known constitutive relations [20]:

$$\mathbf{D} = \varepsilon_0 \varepsilon \mathbf{E} + c^{-1}(\kappa + i\chi)\mathbf{H}, \qquad \mathbf{B} = \mu_0 \mu \mathbf{H} + c^{-1}(\kappa - i\chi)\mathbf{E}, \qquad (1)$$

where $\varepsilon$ is the relative permittivity, $\mu$ is the relative permeability, $\varepsilon_0$ and $\mu_0$ are the vacuum permittivity and permeability, and $c = 1/\sqrt{\varepsilon_0 \mu_0}$ is the speed of light in vacuum. The plane waves in an unbounded bi-isotropic medium can be decomposed into two independent wave field components corresponding to two circularly polarized electromagnetic waves. Specifically, the electric field and the magnetic fields **E** and **H** can be written in the forms $\mathbf{E} = \mathbf{E}_+ + \mathbf{E}_-$ and $\mathbf{H} = \mathbf{H}_+ + \mathbf{H}_-$, respectively, where the "+" ("-") signal is associated with a right (left) circularly polarized wave (RCP/LCP). The corresponding wave numbers are $k_{+,-} = k_0 n_{+,-}$, where $n_{+,-} = \sqrt{\varepsilon\mu - \kappa^2} \pm \chi$ are the refractive indices. The chirality and Tellegen parameters, $\chi$ and $\kappa$, are responsible for the magnetoelectric coupling of bi-isotropic media.

### A. The scattering problem

We consider first that a plane wave propagates in the air region (e.g. $z < 0$) and illuminates the stratified structure. The transverse electric and magnetic fields that propagate in a generic Tellegen or dielectric medium slab are given by (the variation of the fields along the *x* and *y* directions, if any, is omitted)

$$\mathbf{E}^\pm(z) = e^{\pm i\beta z}\begin{pmatrix} E_x^\pm & E_y^\pm \end{pmatrix}^\mathrm{T}, \qquad \mathbf{H}^\pm(z) = \pm e^{\pm i\beta z}\, \bar{\mathbf{Y}}^\pm \cdot \mathbf{E}^\pm, \qquad (2)$$

where "T" represents the transpose of the vector, and the superscripts $\pm$ indicate if the wave propagates along the $+$ or $-$ *z*-direction. The characteristic impedance matrices are defined such that $\mathbf{E}^+(z) = \bar{\mathbf{Z}}^+ \cdot \mathbf{H}^+(z)$ and $\mathbf{E}^- = -\bar{\mathbf{Z}}^- \cdot \mathbf{H}^-$ and are given by

$$\bar{\mathbf{Z}}^\pm = \frac{\mu_0 \mu \omega}{(n^2+\kappa^2)k_0^2\beta}\begin{bmatrix} k_x k_y \mp \kappa k_0 \beta & k_y^2 + \beta^2 \\ -k_x^2 - \beta^2 & -k_x k_y \mp \kappa k_0 \beta \end{bmatrix}. \qquad (3)$$

The admittance matrices are the inverse of the impedance matrices $\bar{\mathbf{Y}}^+ = (\bar{\mathbf{Z}}^+)^{-1}$ and $\bar{\mathbf{Y}}^- = (\bar{\mathbf{Z}}^-)^{-1}$. The longitudinal propagation constant in the pertinent slab is $\beta = \sqrt{k_0^2 n^2 - k_x^2 - k_y^2}$ where $n = \sqrt{\varepsilon\mu - \kappa^2}$ is the refractive index, $k_0 = \omega/c$, and

$(k_x, k_y)$ are the transverse wave numbers determined by the incidence angle. For a conventional dielectric one should use $\kappa = 0$ in Eq. (3).

On the other hand, for a chiral slab the propagation constants depend on the wave polarization state and thus the formulas for the transverse fields are more intricate. To avoid complicating excessively the formalism, we only show the formulas for the particular case of normal incidence ($k_x = k_y = 0$), which is the focus of our study. The fields in a chiral slab for $k_x = k_y = 0$ can be written as:

$$\begin{aligned} \mathbf{E}^+(z) &= e^{i\beta_+ z} E_P^+ \mathbf{e}_+ + e^{i\beta_- z} E_M^+ \mathbf{e}_- & \mathbf{E}^-(z) &= e^{-i\beta_- z} E_P^- \mathbf{e}_+ + e^{-i\beta_+ z} E_M^- \mathbf{e}_- \\ \mathbf{H}^+(z) &= -\frac{i}{\eta}\left(e^{i\beta_+ z} E_P^+ \mathbf{e}_+ - e^{i\beta_- z} E_M^+ \mathbf{e}_-\right), & \mathbf{H}^-(z) &= \frac{i}{\eta}\left(e^{-i\beta_- z} E_P^- \mathbf{e}_+ - e^{-i\beta_+ z} E_M^- \mathbf{e}_-\right) \end{aligned} \quad (4)$$

where $\eta = \sqrt{\mu_0 \mu / \varepsilon_0 \varepsilon}$ is the chiral medium wave impedance. For a chiral slab the longitudinal wave numbers satisfy $\beta_\pm = k_0 n_\pm$ with the refractive indices $n_\pm = \sqrt{\varepsilon \mu} \pm \chi$. A chiral medium is thus a birefringent medium with two distinct refractive indices. For propagation along the +z-direction the normalized wave fields may be taken equal to $\mathbf{e}_+ = \hat{\mathbf{x}} + i\hat{\mathbf{y}}$ and $\mathbf{e}_- = \hat{\mathbf{x}} - i\hat{\mathbf{y}}$, and propagate with the exponential factors $e^{i\beta_+ z}$ and $e^{i\beta_- z}$, respectively. Note that for propagation along the –z direction the plane waves associated with the same eigenstates $\mathbf{e}_+$ and $\mathbf{e}_-$ have the propagation factors $e^{-i\beta_- z}$ and $e^{-i\beta_+ z}$, respectively.

The complex valued field amplitudes in each slab $E_P^\pm, E_M^\pm$ are linked at the interfaces ($z = z_i$) through the usual continuity equations:

$$\begin{aligned} \mathbf{E}_i^+(z_i) + \mathbf{E}_i^-(z_i) &= \mathbf{E}_{i+1}^+(z_i) + \mathbf{E}_{i+1}^-(z_i) \\ \mathbf{H}_i^+(z_i) + \mathbf{H}_i^-(z_i) &= \mathbf{H}_{i+1}^+(z_i) + \mathbf{H}_{i+1}^-(z_i) \end{aligned}, \quad i = 0, \ldots, N \quad (5)$$

where $z_0 = 0$, $z_1 = d_1$, $z_2 = d_1 + d_2$, etc. Thus, if the amplitudes of the incident waves are known ($\mathbf{E}_0^+$ in case of incidence from the left, and $\mathbf{E}_{N+1}^-$ in case of incidence from the right), the remaining field amplitudes can be easily determined by solving a standard linear system. For incidence from the left (with $\mathbf{E}_{N+1}^- = 0$), we define the reflection and transmission matrices, $\bar{\mathbf{R}}$ and $\bar{\mathbf{T}}$, through the formulas $\bar{\mathbf{E}}_0^-(z = z_0) = \bar{\mathbf{R}} \cdot \bar{\mathbf{E}}_0^+(z = z_0)$ and $\bar{\mathbf{E}}_{N+1}^+(z = z_N) = \bar{\mathbf{T}} \cdot \bar{\mathbf{E}}_0^+(z = z_0)$. The $\bar{\mathbf{R}}$ and $\bar{\mathbf{T}}$ matrices may be written explicitly as

$$\bar{\mathbf{R}} = \begin{pmatrix} R_{11} & R_{12} \\ R_{21} & R_{22} \end{pmatrix}, \quad \bar{\mathbf{T}} = \begin{pmatrix} T_{11} & T_{12} \\ T_{21} & T_{22} \end{pmatrix}, \tag{6}$$

where the coefficients $R_{ij}$ and $T_{ij}$ are complex valued numbers.

## B. Scattering matrix

The wave scattering in stratified media can be conveniently studied using the theory of microwave networks [38]. In this context, each channel of propagation is associated with a port. For normal incidence there are two physical channels, one associated with the propagation in the left-hand side air region and another with the propagation in the right-hand side air region (Fig. 1a). However, because there are two allowed polarization states the system is equivalent to a four-port network, as shown in Fig. 1b.

The scattering matrix provides a complete description of the network response and relates the incident and scattered waves, $\mathbf{E}_{inc}$ and $\mathbf{E}_{scatt}$, respectively, as $\mathbf{E}_{scatt} = \bar{\mathbf{S}}_{scatt} \cdot \mathbf{E}_{inc}$. In our case we have $\mathbf{E}_{inc} = \begin{pmatrix} E_{0,x}^+ & E_{0,y}^+ & E_{N+1,x}^- & E_{N+1,y}^- \end{pmatrix}^T$ and $\mathbf{E}_{scatt} = \begin{pmatrix} E_{0,x}^- & E_{0,y}^- & E_{N+1,x}^+ & E_{N+1,y}^+ \end{pmatrix}^T$, where the transverse fields

$\overline{\mathbf{E}}_0^\pm (z = z_0)$, $\overline{\mathbf{E}}_{N+1}^\pm (z = z_N)$ are defined as in the previous subsection. Thus, the scattering (four-by-four) matrix for normal incidence is such that:

$$\overline{\mathbf{S}}_{scatt} = \begin{pmatrix} \overline{\mathbf{R}}^L & \overline{\mathbf{T}}^R \\ \overline{\mathbf{T}}^L & \overline{\mathbf{R}}^R \end{pmatrix}, \qquad (7)$$

where $(\overline{\mathbf{R}}^L, \overline{\mathbf{T}}^L)$ are the reflection and the transmission matrices for an incident wave propagating from the left to the right (L-R). Similarly, $(\overline{\mathbf{R}}^R, \overline{\mathbf{T}}^R)$ are the reflection and the transmission matrices for an incident wave propagating from the right to the left (R-L). The superscripts "L" and "R" indicate from which side (left or right) the incident wave comes from. The matrices $(\overline{\mathbf{R}}^L, \overline{\mathbf{T}}^L)$ are calculated as detailed in the previous subsection and $(\overline{\mathbf{R}}^R, \overline{\mathbf{T}}^R)$ can be found in a similar manner.

A microwave network is reciprocal if and only if the corresponding scattering matrix satisfies [38]:

$$\overline{\mathbf{S}}_{scatt} = (\overline{\mathbf{S}}_{scatt})^T, \qquad (8)$$

where $(\overline{\mathbf{S}}_{scatt})^T$ is the transpose of $\overline{\mathbf{S}}_{scatt}$. For lossless media, the concept of nonreciprocity is strictly linked with the breaking of the time-reversal symmetry in electrodynamics. Thus, the network associated with the scattering matrix defined in Eq. (7) is reciprocal if

$$\overline{\mathbf{R}}^L = (\overline{\mathbf{R}}^L)^T, \quad \overline{\mathbf{R}}^R = (\overline{\mathbf{R}}^R)^T, \quad \overline{\mathbf{T}}^L = (\overline{\mathbf{T}}^R)^T. \qquad (9)$$

It is useful to note that because all the slabs are isotropic and invariant to rotations about the *z*-axis the response of the structure must remain qualitatively the same, independent of the direction of the incoming electric field. This implies that for normal incidence all the reflection and transmission dyadics $(\overline{\mathbf{T}}^L, \overline{\mathbf{T}}^R, \overline{\mathbf{R}}^L, \overline{\mathbf{R}}^R)$ are necessarily of the type

$$\overline{\mathbf{A}} = \begin{pmatrix} a_{11} & a_{12} \\ -a_{12} & a_{11} \end{pmatrix}. \tag{10}$$

Thus, the diagonal elements must be equal and the anti-diagonal elements must be additive inverses. For future reference, it is noted that that the sum of two matrices $\overline{\mathbf{A}}_a, \overline{\mathbf{A}}_b$ of the form defined in Eq. (10) is still a matrix of the same type. The inverse of $\overline{\mathbf{A}}$ is also a matrix of the same type. Moreover, two matrices $\overline{\mathbf{A}}_a, \overline{\mathbf{A}}_b$ of the form presented in Eq. (10) commute, $\overline{\mathbf{A}}_a \cdot \overline{\mathbf{A}}_b = \overline{\mathbf{A}}_b \cdot \overline{\mathbf{A}}_a$, and the product is still a matrix of the same type.

### C. Eigenstates of the scattering problem

The eigenstates of the scattering problem illustrated in Fig. 1 are circularly polarized waves. To show this we note that the eigenvectors of a matrix $\overline{\mathbf{A}}$ of the type presented in Eq. (10) are $\mathbf{e}_+ = \begin{pmatrix} 1 & i \end{pmatrix}^\mathrm{T}$, $\mathbf{e}_- = \begin{pmatrix} 1 & -i \end{pmatrix}^\mathrm{T}$, such that:

$$\overline{\mathbf{A}} \cdot \mathbf{e}_+ = a_+ \mathbf{e}_+, \qquad \overline{\mathbf{A}} \cdot \mathbf{e}_- = a_- \mathbf{e}_-, \tag{11}$$

with $a_+ = a_{11} + i a_{12}$ and $a_- = a_{11} - i a_{12}$. Because the reflection and transmission matrices are of the form defined in Eq. (10), it follows that when the incident field is $\mathbf{E}_{inc} = E_{inc} \mathbf{e}_\pm$ then the corresponding reflected and transmitted waves are $\mathbf{E}_{ref} = E_{inc} R_\pm \mathbf{e}_\pm$ and $\mathbf{E}_{tx} = E_{inc} T_\pm \mathbf{e}_\pm$, respectively, where $T_\pm = T_{11} \pm i T_{12}$ and $R_\pm = R_{11} \pm i R_{12}$. The polarization states $\mathbf{e}_\pm$ are evidently associated with circularly polarized waves. Hence, when $\mathbf{E}_{inc} = E_{inc} \mathbf{e}_\pm$, it is seen that the transmitted wave has always the same polarization state as the incident wave, whereas the reflected wave is also circularly polarized but has handedness opposite to that of the incoming wave.

When the incoming wave propagates from the left to the right (L-R) the pertinent transmission and reflection matrices are $(\overline{\mathbf{T}}^L, \overline{\mathbf{R}}^L)$ and the corresponding eigenvalues are $T_{\pm}^L = T_{11}^L \pm iT_{12}^L$ and $R_{\pm}^L = R_{11}^L \pm iR_{12}^L$. Similarly, for propagation from the right to the left (R-L) the relevant transmission and reflection matrices are $\left(\overline{\mathbf{T}}^R, \overline{\mathbf{R}}^R\right)$ and the corresponding eigenvalues are $T_{\pm}^R = T_{11}^R \pm iT_{12}^R$ and $R_{\pm}^R = R_{11}^R \pm iR_{12}^R$. Note that $\mathbf{e}_+$ and $\mathbf{e}_-$ correspond to right and left circularly polarized waves (RCP and LCP) for a wave propagating along the $+z$ direction. On the other hand, for a wave propagating along the $-z$ direction, $\mathbf{e}_+$ and $\mathbf{e}_-$ are associated with LCP and RCP waves, respectively. In summary, an incident RCP/LCP wave illuminating the stratified structure originates a reflected LCP/RCP wave and a transmitted RCP/LCP wave.

The reciprocity conditions in Eq. (9) can be expressed in terms of $T_{\pm}^R$, $T_{\pm}^L$ and $R_{\pm}^L$, $R_{\pm}^R$. It is straightforward to show that the conditions in Eq. (9) are equivalent to:

$$R_+^L = R_-^L, \qquad R_+^R = R_-^R, \qquad T_{\pm}^L = T_{\mp}^R. \tag{12}$$

In other words, for reciprocal systems the reflection coefficients for LCP and RCP waves must be the same for incidences from the same side of the structure. On the other hand, the transmission coefficient is required to be same for waves with the same handedness, and incidence from opposite sides of the structure. Thus, as illustrated in Fig. 2, incident RCP/LCP waves coming from the L-R and R-L directions must have the same transmission coefficients in case of stratified structures formed by conventional dielectrics and/or chiral media.

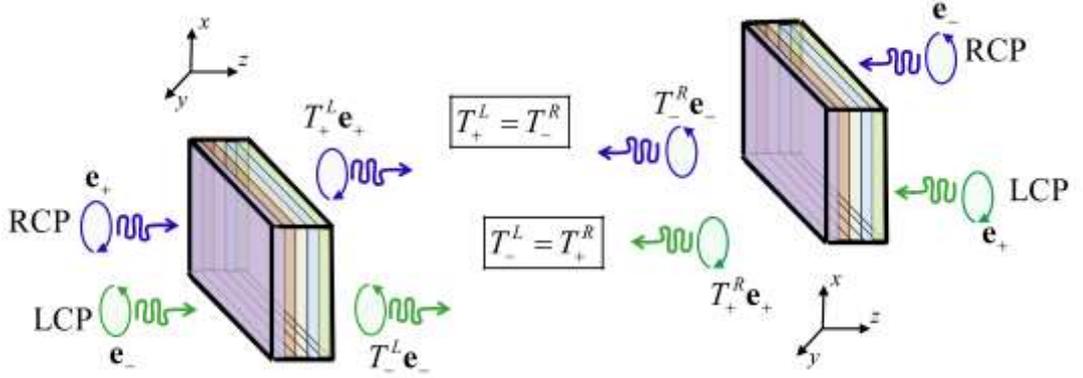

**Fig. 2** (Color online) Symmetric transmission of circularly polarized waves incident on the front and back sides of planar structures made of reciprocal materials, such as, conventional dielectrics and/or chiral media.

### D. Stratified Tellegen media

Next, we consider the particular case wherein all the slabs are either conventional dielectrics or Tellegen materials. In Appendix B, it is proven that, independent of the number of layers of a stratified Tellegen structure, the transmission matrix for incidence L-R is exactly the same as the transmission matrix for incidence R-L:

$$\overline{\mathbf{T}}^L = \overline{\mathbf{T}}^R. \tag{13}$$

In particular, the transmission coefficients for circularly polarized waves satisfy $T_+ \equiv T_+^R = T_+^L$ and $T_- \equiv T_-^R = T_-^L$.

To understand the role of the nonreciprocal response of Tellegen media in the wave scattering, next we present a few illustrative numerical examples. First, we consider the case wherein a single Tellegen slab stands alone in free-space (this corresponds to $d_2 = ... = d_N = 0$ in Fig. 1a). In Fig. 3 we depict the reflection and transmission coefficients for an incoming circularly polarized wave for the constitutive parameters $\varepsilon = 2$, $\mu = 1$, $\kappa = 0.1$ and thickness $d$. The absolute values of the transmission and reflection coefficients for a wave propagating in the directions L-R and R-L, $\left(R_+^L, R_-^L\right)$, $\left(R_+^R, R_-^R\right)$, $T_+ \equiv T_+^R = T_+^L$ and $T_- \equiv T_-^R = T_-^L$, are depicted in

Fig. 3a. The phases of the same scattering parameters are shown in Fig. 3b. Consistent with the fact that the time-reversal symmetry is broken in Tellegen media, it is found that $R_+^L \neq R_-^L$ and $R_+^R \neq R_-^R$ [see Eq. (12)]. Thus, the nonreciprocal response of the structure may be detected by inspection of the reflection coefficients. As seen in Fig. 3, we have $T_+ = T_-$ and thus, despite the nonreciprocal response of the Tellegen material, the reciprocity criterion $T_\pm^L = T_\mp^R$ [see Eq. (12)] is still satisfied. In this example, the property $T_\pm^L = T_\mp^R$ is a simple consequence of the "left-right" symmetry of the structure. Indeed, we numerically verified that the condition $T_\pm^L = T_\mp^R$ is always observed for spatially symmetric structures of the type Air-$T_1$-$T_2$-$T_3$-(…)-$T_N$-(…)-$T_3$-$T_2$-$T_1$-Air, where "$T_i$" corresponds to some Tellegen medium.

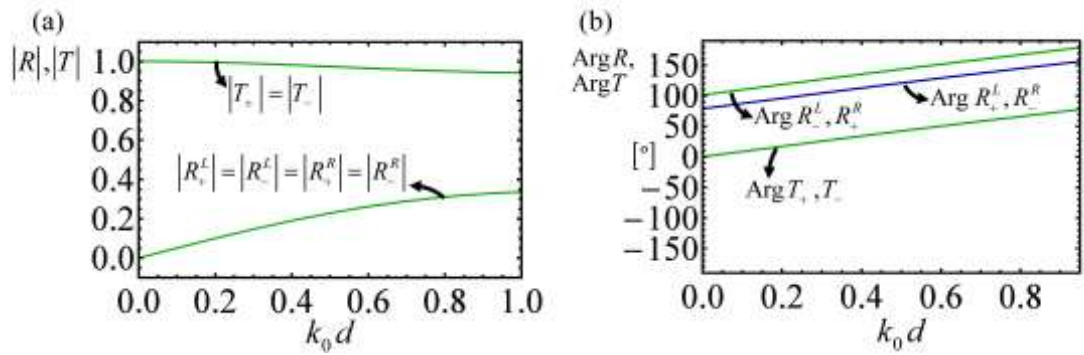

**Fig. 3** (Color online) Reflection and transmission coefficients for an incoming wave propagating in the L-R and R-L directions for a Tellegen slab, with thickness $d$, standing in vacuum. (a) Amplitude of the reflection and transmission coefficients $R_+^L, R_-^L$ , $R_+^R, R_-^R$ and $T_+, T_-$ . (b) Phase of the reflection and transmission coefficients $R_+^L, R_-^L$ , $R_+^R, R_-^R$ and $T_+, T_-$ .

We also found out that generally one can have $T_+ = T_- = T$ even for asymmetric stratified structures (formed by several slabs), provided each and every subset of three materials (including the air regions) have linear dependent constitutive parameters, such that:

$$\det \begin{pmatrix} \varepsilon_1 & \mu_1 & \kappa_1 \\ \varepsilon_2 & \mu_2 & \kappa_2 \\ \varepsilon_3 & \mu_3 & \kappa_3 \end{pmatrix} = 0. \tag{14}$$

In such conditions, we say that all the materials are in the same *equivalence class* [39]. It was shown by us in Ref. [39] that in this case the structure is reducible under a suitable duality transformation to a stratified structure formed by conventional dielectrics with $\kappa = 0$, i.e. the Tellegen parameter can be formally eliminated. In Appendix C, we present a brief overview of duality transformations. Thus, because the transmission matrix $\bar{\mathbf{T}}$ in the transformed problem reduces to a scalar, it follows that it also reduces to the same scalar in the original structure, and hence $T_{12} = 0$ and thus $T_+ = T_- = T$. Note that this argument only works for the transformed transmission matrix and it does not apply to the transformed reflection matrix. The reason is that the transmission coefficient for the electric field (defined as $\mathbf{E}_T = T \cdot \mathbf{E}_{inc}$) is the same as the transmission coefficient for the magnetic field (defined as $\mathbf{H}_T = T \cdot \mathbf{H}_{inc}$), and thus the transmitted electric and magnetic fields are transformed in the same manner in a scattering problem involving only conventional dielectrics. On the other hand, the reflection coefficients for the transverse electric and magnetic fields are the symmetric of one another. This difference is of crucial importance because a duality transformation mixes the electric and magnetic fields (see Appendix C).

In the second example [see Fig. 4], we characterize the wave scattering by two juxtaposed slabs with thicknesses $d_1$ and $d_2$ embedded in vacuum. The parameters of the Tellegen media and the parameters of the vacuum are chosen such that the structure is not reducible to conventional dielectrics, i.e., the material parameters do not satisfy Eq. (14). Because of this, it is possible to simultaneously break all the criteria in Eq. (12) and detect the nonreciprocal response of the involved materials both in transmission and reflection. The simplest structure that can reveal these effects corresponds to a Tellegen slab juxtaposed to a dielectric slab. This is illustrated in Fig. 4a for the case $d_1 = d_2 = 0.5d$ and for materials with the constitutive parameters $\varepsilon_1 = 2$, $\mu_1 = 1$, $\kappa_1 = 0.1$ and $\varepsilon_2 = 2$, $\mu_2 = 1$. Figure 4b shows that by juxtaposing two Tellegen slabs with constitutive parameters $\varepsilon_1 = 2$, $\mu_1 = 1$, $\kappa_1 = 0.1$ and $\varepsilon_2 = 2.5$, $\mu_2 = 1.5$, $\kappa_2 = 0.5$, and with $d_1 = d_2 = 0.5d$ it is possible to enhance the nonreciprocal effects. Finally, Fig. 4c shows how the strength of the Tellegen parameter $\kappa_1$ determines the asymmetric response.

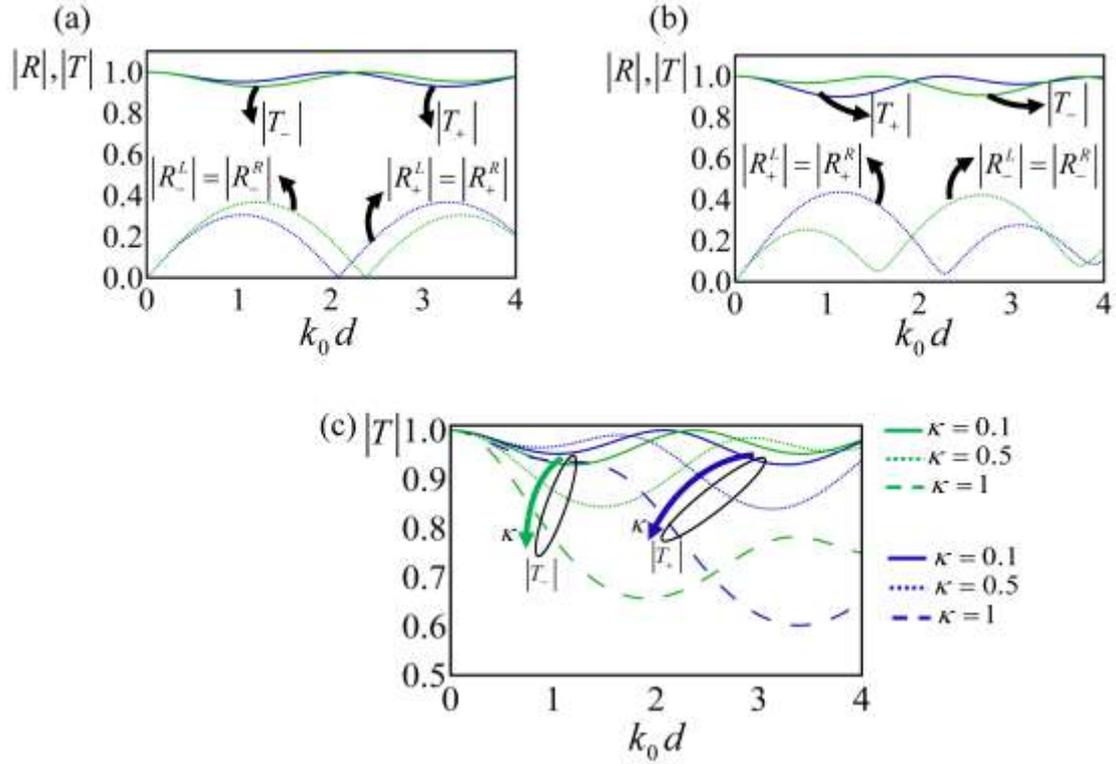

**Fig. 4** (Color online) Amplitude of reflection and transmission coefficients for an incoming wave propagating in the L-R and R-L directions for stratified structures with Tellegen media. The total thickness is $d = d_1 + d_2$ with $d_1 = d_2$. (a) Amplitude of the reflection and transmission coefficients $R_+^L, R_-^L, R_+^R, R_-^R$ and $T_+, T_-$ for a Tellegen slab juxtaposed to a dielectric slab. (b) Amplitude of the reflection and transmission coefficients $R_+^L, R_-^L, R_+^R, R_-^R$ and $T_+, T_-$ for two juxtaposed Tellegen slabs.

(c) Transmission coefficients $T_+, T_-$ for the example (a), but for different values of $\kappa_1$.

## III. Stratified Structures with Ferrites

It is interesting to compare the scattering provided by media with a spontaneous nonreciprocal response, with the more conventional ferrites, which as previously discussed require an external biasing and thus have an anisotropic response. A ferrite biased with a static magnetic field oriented along the $z$-direction is characterized by the following constitutive relations [38]

$$\mathbf{D} = \varepsilon_0 \varepsilon_f \mathbf{E}, \quad \mathbf{B} = \mu_0 \bar{\mu}_f \cdot \mathbf{H}, \tag{15}$$

where $\varepsilon_f$ is the relative permittivity and $\bar{\mu}_f$ is the relative permeability matrix given by

$$\bar{\mu}_f = \begin{pmatrix} \mu_f & -i\kappa_f & 0 \\ i\kappa_f & \mu_f & 0 \\ 0 & 0 & 1 \end{pmatrix}. \tag{16}$$

In case of negligible material loss, the parameters $\mu_f, \kappa_f$ depend on frequency as

$$\mu_f = 1 + \frac{\omega_R \omega_m}{\omega_R^2 - \omega^2}, \quad \kappa_f = \frac{\omega \omega_m}{\omega_R^2 - \omega^2}, \tag{17}$$

where $\omega_R$ is the precession or Larmor frequency and $\omega_m$ is the is the electron Larmor frequency at the saturation magnetization of the ferrite [38].

Here, we are interested in structures with a geometry analogous to that of Fig. 1a, but we allow some of the slabs to be ferrites biased along the $z$-direction. For propagation along the stratification direction the transverse electric and magnetic fields in a generic ferrite slab can be written as a superposition of circularly polarized waves propagating along the positive and negative $z$-directions:

$$\mathbf{E}^+(z) = e^{i\beta_+ z} E_P^+ \mathbf{e}_+ + e^{i\beta_- z} E_M^+ \mathbf{e}_-, \qquad \mathbf{H}^+(z) = \frac{-i}{\eta_+} e^{i\beta_+ z} E_P^+ \mathbf{e}_+ + \frac{i}{\eta_-} e^{i\beta_- z} E_M^+ \mathbf{e}_-$$

$$\mathbf{E}^-(z) = e^{-i\beta_+ z} E_P^- \mathbf{e}_+ + e^{-i\beta_- z} E_M^- \mathbf{e}_-, \qquad \mathbf{H}^-(z) = -\left(\frac{-i}{\eta_+} e^{-i\beta_+ z} E_P^- \mathbf{e}_+ + \frac{i}{\eta_-} e^{-i\beta_- z} E_M^- \mathbf{e}_-\right) \quad (18)$$

where the longitudinal wavenumbers satisfy $\beta_\pm = k_0 n_\pm$ with the refractive indices written as $n_\pm = \sqrt{\varepsilon_f \mu_\pm}$ with $\mu_\pm = \mu_f \pm \kappa_f$ and $\eta_\pm = \sqrt{\mu_0 \mu_\pm / \varepsilon_0 \varepsilon_f}$. Similar to the previous section, the reflection and transmission matrices can be found from the solution of a linear system, and the eigenstates of the scattering problem are circularly polarized waves.

In Fig. 5 we show the absolute values of the reflection and transmission coefficients for a single ferrite slab embedded in a vacuum. As seen, the nonreciprocity is detected in both the reflection and transmission coefficients because $|R_+^L| \neq |R_-^L|$, $|R_+^R| \neq |R_-^R|$, $|T_+^L| \neq |T_-^R|$ and $|T_+^R| \neq |T_-^L|$. In our simulations it was assumed that $\omega_R d/c = 4$ and $\omega_m d/c = 2$, and, for simplicity, the relative permittivity was taken equal to $\varepsilon_f = 1$.

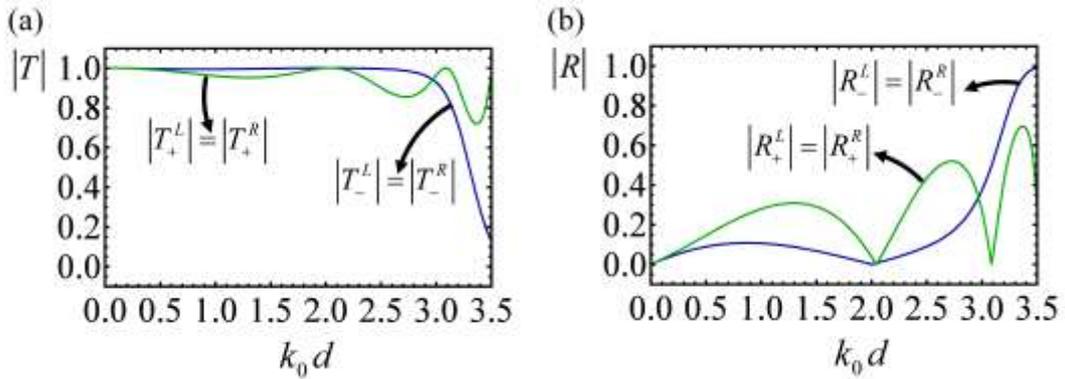

**Fig. 5** (Color online) Amplitude of the transmission and reflection coefficients for a single ferrite slab with thickness $d$. (a) Amplitude of the transmission coefficients, $T_+^L, T_-^L, T_+^R, T_-^R$, for a wave propagating in the L-R and R-L directions, respectively. (b) Amplitude of the reflection coefficients, $R_+^L, R_-^L, R_+^R, R_-^R$, for a wave propagating in the L-R and R-L directions, respectively.

Hence the nonreciprocity of a single ferrite slab is manifested in both the reflected and transmitted waves, different from what happens in stratified Tellegen structures with media in the same equivalence class [see Sect. II.D], and in particular different from a single Tellegen slab. Thus, we conclude that to mimic the behaviour of a single ferrite slab, we need at least *two* juxtaposed Tellegen slabs (or alternatively a Tellegen slab and a standard dielectric) in such a way that the two Tellegen media and the vacuum are not in the same equivalence class.

## IV. Photonic Crystals with Nonreciprocal Media

With the aim of further exploring the opportunities created by a spontaneous nonreciprocal response, next we investigate the photonic band diagrams of periodic layered structures formed by Tellegen media or ferrites combined with chiral media and/or conventional dielectrics. It is proven that in some conditions the photonic band diagrams can exhibit an asymmetry such that $\omega(k_z) \neq \omega(-k_z)$. We restrict our attention to Bloch modes with $k_x = k_y = 0$, which corresponds to propagation along the direction of stratification.

### A. Dispersion equations for the Bloch waves

Periodic structures may be seen as a cascade of identical multi-port networks. In particular, in case of normal incidence a stratified (layered) structure [see Fig. 6] is equivalent to a 4-port network, as discussed previously. Thus, each unit cell, independent of its complexity or of the involved materials, is fully characterized by the reflection and transmission matrices for L-R and R-L incidences, i.e., $\left( \bar{\mathbf{R}}^L, \bar{\mathbf{T}}^L \right)$ and $\left( \bar{\mathbf{R}}^R, \bar{\mathbf{T}}^R \right)$ respectively. In particular, the electric fields at the front and back interfaces of the *n*-th interface are related by:

$$\mathbf{E}^n_{scatt} = \bar{\mathbf{R}}^L \cdot \mathbf{E}^n_{inc} + \bar{\mathbf{T}}^R \cdot \mathbf{E}^{n+1}_{inc}, \qquad \mathbf{E}^{n+1}_{scatt} = \bar{\mathbf{T}}^L \cdot \mathbf{E}^n_{inc} + \bar{\mathbf{R}}^R \cdot \mathbf{E}^{n+1}_{inc}. \tag{19}$$

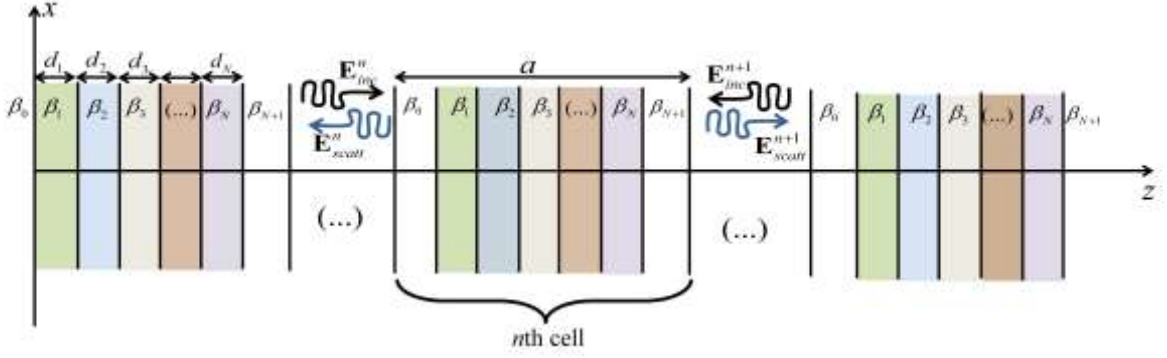

**Fig. 6** (Color online) A periodic layered structure formed by different media.

For Bloch waves it is necessary that $\mathbf{E}^{n+1}_{scatt} = e^{ik_z a} \mathbf{E}^n_{inc}$ and $\mathbf{E}^{n+1}_{inc} = e^{ik_z a} \mathbf{E}^n_{scatt}$, where $k_z$ is the propagation constant of the periodic structure and $a$ is the spatial period. Substituting this result into Eq. (19) one obtains after simple manipulations

$$\begin{bmatrix} e^{-ik_z a}\bar{\mathbf{T}}^L - \bar{\mathbf{I}}_t & \bar{\mathbf{R}}^R \\ \bar{\mathbf{R}}^L & e^{ik_z a}\bar{\mathbf{T}}^R - \bar{\mathbf{I}}_t \end{bmatrix} \cdot \begin{bmatrix} \mathbf{E}^n_{inc} \\ \mathbf{E}^n_{scatt} \end{bmatrix} = \begin{bmatrix} 0 \\ 0 \end{bmatrix}, \tag{20}$$

where $\bar{\mathbf{I}}_t$ is the identity 2×2 dyadic. To have a nontrivial solution, the determinant of the above matrix must vanish [2]:

$$\det\left(\begin{bmatrix} e^{-ik_z a}\bar{\mathbf{T}}^L - \bar{\mathbf{I}}_t & \bar{\mathbf{R}}^R \\ \bar{\mathbf{R}}^L & e^{ik_z a}\bar{\mathbf{T}}^R - \bar{\mathbf{I}}_t \end{bmatrix}\right) = 0. \tag{21}$$

To make further progress, we note that because the polarization eigenstates of our stratified structures are circularly polarized, the Bloch eigenstates must be such that $\begin{bmatrix} \mathbf{E}^n_{inc} \\ \mathbf{E}^n_{scatt} \end{bmatrix} = \begin{bmatrix} E^n_{inc}\mathbf{e}_\pm \\ E^n_{scatt}\mathbf{e}_\pm \end{bmatrix}$ with $\mathbf{e}^+ = \begin{pmatrix} 1 & i \end{pmatrix}^T$ and $\mathbf{e}^- = \begin{pmatrix} 1 & -i \end{pmatrix}^T$. Thus, from Eq. (20) we can factorize the dispersion equation into two independent equations:

$$\left(T^L_+ e^{-ik_z a} - 1\right)\left(T^R_+ e^{ik_z a} - 1\right) - R^L_+ R^R_+ = 0, \qquad \left(T^L_- e^{-ik_z a} - 1\right)\left(T^R_- e^{ik_z a} - 1\right) - R^L_- R^R_- = 0. \tag{22}$$

Note that the reflection and transmission coefficients are functions of $\omega$.

For periodic structures formed by conventional dielectrics one has $R_+^L = R_-^L$, $R_+^R = R_-^R$, $T_+^R = T_-^L = T_+^L = T_-^R$, and thus the two dispersion equations are coincident. In that case, the Bloch eigenmodes are degenerate. However, both for chiral and Tellegen media in general the two equations yield modes with different dispersions. The modes associated with the "+" ("-") sign correspond to a superposition of RCP (LCP) waves propagating in the +z-direction and LCP (RCP) waves propagating in the -z-direction. It is easy to check using Eq. (12) that for reciprocal materials (e.g. chiral media) the dispersion of the modes associated with "+" sign, $\omega_+(k_z)$, and the dispersion of the modes associated with "–" sign, $\omega_-(k_z)$, are such that $\omega_+(k_z) = \omega_-(-k_z)$. Thus, as could be expected, for reciprocal materials the photonic band diagrams are always symmetric.

### B. Band diagrams for Tellegen periodic structures

Let us now consider the particular case wherein all the materials are either Tellegen media or conventional dielectrics. It was proven in Sect. II.D that in this situation $\overline{\mathbf{T}}^L = \overline{\mathbf{T}}^R$, and thus we can put $T_+ \equiv T_+^R = T_+^L$ and $T_- \equiv T_-^R = T_-^L$. Thus, the characteristic equations defined in Eq. (22) reduce to

$$\cos(k_z a) = \frac{T_+}{2} + \frac{1}{2T_+}\left(1 - R_+^L R_+^R\right), \quad \cos(k_z a) = \frac{T_-}{2} + \frac{1}{2T_-}\left(1 - R_-^L R_-^R\right). \tag{23}$$

In particular, it is evident that $\omega_\pm(k_z) = \omega_\pm(-k_z)$ and hence, despite the nonreciprocal response of Tellegen media, the band diagrams exhibit always spectral symmetry.

In general, the Bloch modes are not degenerate and $\omega_+ \neq \omega_-$. However, when all the materials in the unit cell are in the same equivalence class, the photonic crystal is reducible under some duality transformation to a photonic crystal containing only conventional dielectric media. One important property of duality transformations is that they only act over the electromagnetic fields, leaving the spatial coordinates $\mathbf{r} = (x, y, z)$ and the time coordinate $t$ invariant (see Appendix C). As a consequence, the band structure of a photonic crystal is invariant under a duality mapping, i.e the dispersion diagrams $\omega$ vs. $\mathbf{k}$ are precisely the same for the original crystal and for a duality-transformed crystal [39]. This shows that the Bloch modes of a Tellegen photonic crystal with media in the same equivalence class must be degenerate $\omega_+ = \omega_-$ for propagation along the stratification direction.

To illustrate the discussion, we consider a photonic crystal such that the unit cell is formed by an air slab with thickness $d_0 = 0.5a$ and a Tellegen slab with thickness $d_1 = 0.5a$. The constitutive parameters of the Tellegen medium are given by $\varepsilon_1 = 2$, $\mu_1 = 1$ and $\kappa_1 = 0.8$. The computed band structure is depicted in Fig. 7a. As seen, because any two Tellegen materials define an equivalence class the dispersions of the modes "+" and "−" are exactly coincident.

In the second example, we consider a crystal such that the unit cell is formed by three material slabs: an air slab with thickness $d_0 = 0.2a$, a Tellegen slab and a dielectric slab with thicknesses $d_1 = 0.4a$ and $d_2 = 0.4a$. The constitutive parameters of the two media are $\varepsilon_1 = 2$, $\mu_1 = 1$ and $\kappa_1 = 0.8$ and $\varepsilon_2 = 2$, $\mu_2 = 1$, respectively.

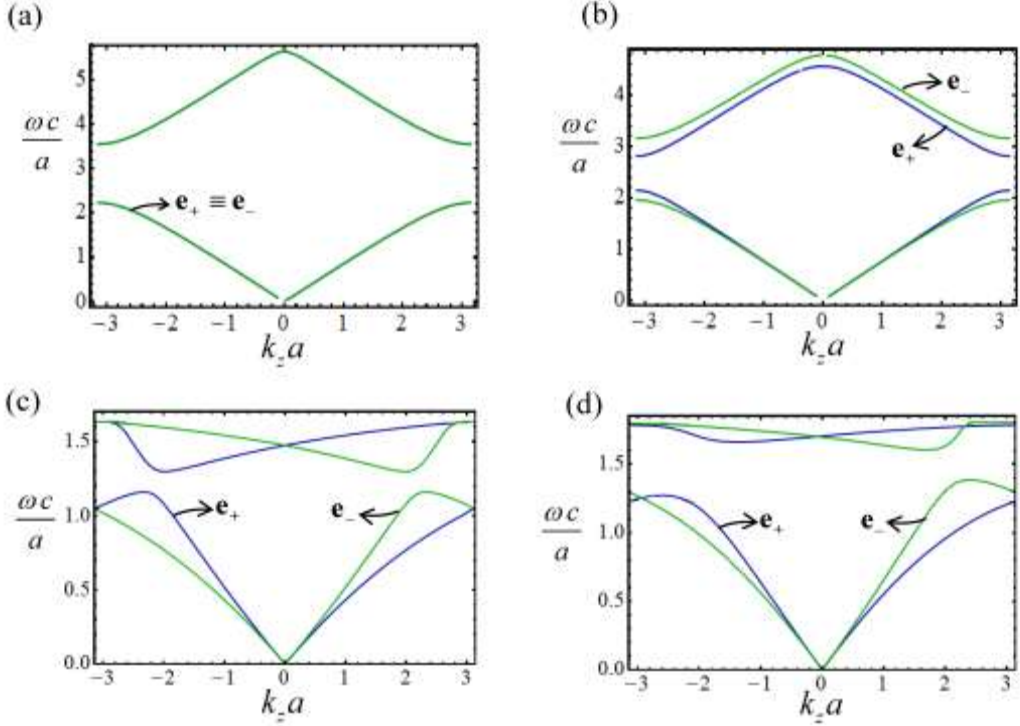

**Fig. 7** (Color online) Band diagrams of periodic structures formed by bi-isotropic media. The lattice constant is $a$. (a) Unit cell with Tellegen media in the same class. (b) Unit cell with Tellegen media in different classes. (c) Unit cell with a dielectric medium and a chiral medium. (d) Unit cell with a dielectric medium, a chiral medium and a Tellegen medium.

It is clear from Fig. 7b that for this photonic crystal the band diagrams corresponding to the two polarization eigenstates $\mathbf{e}_+$ and $\mathbf{e}_-$ are different. This happens because the considered media are not simultaneously reducible to conventional dielectrics.

For completeness, we mention that for symmetric unit cells of the type Air-$T_1$-$T_2$-$T_3$-(…)-$T_N$-(…)-$T_3$-$T_2$-$T_1$-Air one also obtains degenerate modes ($\omega_+ = \omega_-$) even when the different media do not belong to the same equivalence class. Indeed, in such a case we have $T_+ = T_-$ [see Sect. II.D] and because of the symmetry of the problem it is evident that $R_\pm^L = R_\mp^R$.

## C. Sufficient conditions for spectral symmetry

The band diagrams $\omega(k_z)$ depicted in the Fig. 7a and Fig. 7b have the spectral symmetry, $\omega(k_z) = \omega(-k_z)$. It is well known that such a property is characteristic of structures that either satisfy the Lorentz reciprocity theorem or that are invariant under a spatial-inversion transformation $\mathbf{r} \to -\mathbf{r}$. However, in general a stratified photonic crystal formed by Tellegen media is not protected by such transformations. Hence, it is quite remarkable that our numerical simulations give $\omega(k_z) = \omega(-k_z)$. In what follows, we prove that such result is not accidental and that the spectral symmetry of the considered photonic crystals is protected (for propagation along the $z$-direction) by another symmetry of the system.

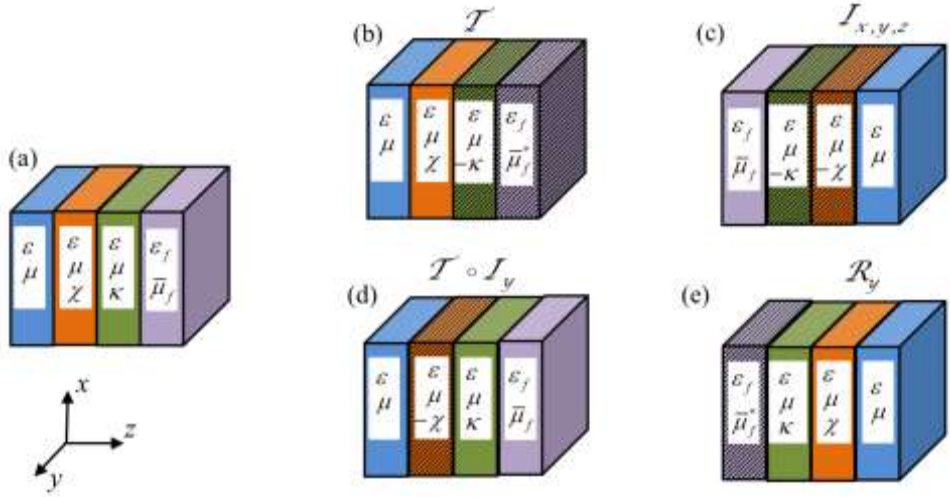

**Fig. 8** (Color online) Transformation theory for stratified periodic crystals with bi-isotropic and ferrite materials. (a) Original photonic crystal. (b)-(e) Structure of the photonic crystal after the electromagnetic fields are subjected to the indicated operation. (b) Time reversal operation, $\mathcal{T}$. (c) Spatial inversion transformation $\mathcal{I}_{x,y,z}: \mathbf{r} \to -\mathbf{r}$. (d) Composition of time reversal and $y$-spatial inversion, $\mathcal{T} \circ \mathcal{I}_y$. (e) Rotation of 180º about the $y$-axis, $\mathcal{R}_y: \mathbf{r} \to (-x, y, -z)$.

Figure 8 illustrates how different symmetry transformations of the electromagnetic fields affect the structure of a stratified photonic crystal. All the transformations map $k_z$ into $-k_z$. Thus, if the photonic crystal stays invariant under one of these transformations then the spectral symmetry is protected by the transformation.

Next, we focus on the action of the transformation sketched in Fig. 8d and prove that the spectral symmetry of the Tellegen photonic crystals is a consequence of this symmetry transformation. Thus the ferrite and the chiral slabs in Fig. 8d should be ignored in the following discussion. Let $\mathbf{D}, \mathbf{E}, \mathbf{B}, \mathbf{H}$ be the electromagnetic fields associated with a Bloch mode characterized by the wave number (for propagation along the z-direction) $k_z$ with the oscillation frequency $\omega$. Let us define the auxiliary fields:

$$\begin{aligned}\mathbf{D}'(\mathbf{r}) &= \mathcal{I}_y \cdot \mathbf{D}^*(\mathcal{I}_y \cdot \mathbf{r}), & \mathbf{E}'(\mathbf{r}) &= \mathcal{I}_y \cdot \mathbf{E}^*(\mathcal{I}_y \cdot \mathbf{r}) \\ \mathbf{B}'(\mathbf{r}) &= \mathcal{I}_y \cdot \mathbf{B}^*(\mathcal{I}_y \cdot \mathbf{r}), & \mathbf{H}'(\mathbf{r}) &= \mathcal{I}_y \cdot \mathbf{H}^*(\mathcal{I}_y \cdot \mathbf{r})\end{aligned}, \quad (24)$$

where $\mathcal{I}_y : (x, y, z) \to (x, -y, z)$ represents an inversion of the y-coordinate. The above transformation is equivalent to a composition of the time-reversal operator and the inversion of the y-coordinate operator. It can be checked that the primed fields satisfy the Maxwell equations (with the same oscillation frequency $\omega$) in a transformed photonic crystal characterized by the material matrix:

$$\bar{\mathbf{M}}'(\mathbf{r}) = \begin{pmatrix} \mathcal{I}_y & 0 \\ 0 & \mathcal{I}_y \end{pmatrix} \cdot \bar{\mathbf{M}}^*(\mathcal{I}_y \cdot \mathbf{r}) \cdot \begin{pmatrix} \mathcal{I}_y & 0 \\ 0 & \mathcal{I}_y \end{pmatrix}. \quad (25)$$

The material matrix $\bar{\mathbf{M}}(\mathbf{r})$ characterizes the original Tellegen photonic crystal and is given by

$$\bar{\mathbf{M}}(\mathbf{r}) = \begin{pmatrix} \varepsilon & \kappa \\ \kappa & \mu \end{pmatrix}. \quad (26)$$

Clearly, because of the conjugation operation, the primed fields define a Bloch wave associated with the wave number $-k_z$. Because our (stratified along z) photonic crystal has parameters independent of the *y* coordinate, and because the parameters of Tellegen media are real-valued, it can be checked that $\bar{\mathbf{M}}'(\mathbf{r}) = \bar{\mathbf{M}}^*(\mathbf{r}) = \bar{\mathbf{M}}(\mathbf{r})$. In other words, a stratified photonic crystal formed by lossless Tellegen media stays invariant under the application of the composition of time-reversal followed by an inversion of the *y*-coordinate (Fig. 8d). Note that the photonic crystal has neither the time reversal symmetry (because Tellegen media are nonreciprocal) nor the *y*-inversion symmetry (because a change in the coordinates $(x, y, z) \to (x, -y, z)$ transforms the Tellegen parameter as $\kappa \to -\kappa$), but it has the symmetry corresponding to the composition of the two operations. This means that the primed fields are Bloch solutions of the Maxwell equations in the original photonic crystal associated with the parameters $(\omega, -k_z)$. This evidently implies that the band structure of the photonic crystal has a spectral symmetry for wave propagation along the *z*-direction, as we wanted to prove.

### D. Band diagrams for periodic structures with bi-isotropic media

The spectral symmetry of photonic crystals with chiral media is not protected by the transformation defined in Eq. (24) because for such structures $\bar{\mathbf{M}}'(\mathbf{r}) = \bar{\mathbf{M}}^*(\mathbf{r}) \neq \bar{\mathbf{M}}(\mathbf{r})$. Indeed, chiral media stay invariant under the time-reversal operation but the chirality parameter is transformed as $\chi \to -\chi$ under the *y*-inversion operation. This suggests that photonic crystals formed by both Tellegen and chiral media may have asymmetric band diagrams. This possibility motivated the study of band diagrams of general photonic crystals formed by bi-isotropic media, which is the topic of this subsection.

To begin with, we consider a photonic crystal with a unit cell formed by a lossless chiral slab and an air layer. In Fig. 7c we depict a representative band diagram and, as expected, it exhibits spectral symmetry because this symmetry is protected by the invariance under time reversal. In our calculations we assumed that the dispersion of the chiral material is described by the Condon model [40]. Within this framework, the constitutive parameters of a lossless chiral medium $\varepsilon(\omega)$, $\mu(\omega)$ and $\chi(\omega)$ depend on the frequency $\omega$ as:

$$\varepsilon(\omega) = \varepsilon_b + \Omega_\varepsilon \omega_R^2 / (\omega_R^2 - \omega^2), \quad \mu(\omega) = \mu_b + \Omega_\mu \omega^2 / (\omega_R^2 - \omega^2),$$
$$\chi(\omega) = \Omega_\chi \omega_R \omega / (\omega_R^2 - \omega^2). \quad (27)$$

In the plot of Fig. 7c the parameters of the Condon model are chosen as: $\varepsilon_b = 1$, $\mu_b = 1.5$, $\omega_R a/c = 2$, $\Omega_\varepsilon = 4$, $\Omega_\mu = 0.5$ and $\Omega_\chi = 1.6$. The thicknesses of the air and chiral regions are $d_0 = 0.5a$ and $d_1 = 0.5a$, respectively.

Next, we consider periodic structures formed by both Tellegen and chiral media. To have a spectral asymmetry we need at least three different material slabs in the unit cell. Indeed any layered photonic crystal formed by uniquely two isotropic slabs in the unit cell stays invariant under a rotation of 180º about the y-axis. Such a transformation does not affect the material parameters of either Tellegen or chiral media (unlike an inversion) (Fig. 8e), and thus it follows that the invariance under an 180º rotation about the y-axis also protects the spectral symmetry.

Therefore, we consider a periodic structure wherein the unit cell is formed by an air slab, a chiral medium slab and a Tellegen medium slab with thicknesses $d_0$, $d_1$ and $d_2$, respectively. The chiral medium has the same constitutive parameters as in the example of Fig. 7c. The Tellegen medium is characterized by the constitutive parameters $\varepsilon_2 = 2$, $\mu_2 = 1$ and $\kappa_2 = 0.8$. The slab thicknesses are $d_0 = 0.25a$, $d_1 = 0.5a$ and $d_2 = 0.25a$. The computed band diagram is shown in Fig. 7d.

Although periodic structures formed by only Tellegen media or by only chiral media exhibit symmetric band diagrams (Fig. 7a, b and c), it is seen that the symmetry is broken in a crystal formed by both chiral and Tellegen media. Indeed, this type of photonic crystals is neither protected by time reversal invariance nor by the transformation defined in Eq. (24).

An immediate consequence of the spectral asymmetry is that the group velocities ($v_g = \partial\omega/\partial k_z$) of the Bloch waves that propagate in the forward and backward directions are not the symmetric of one another. This is illustrated in Fig. 9a for the photonic crystal considered in the Fig. 7d, which represents the group velocities for the two eigenmodes, $\mathbf{e}_+$ and $\mathbf{e}_-$ as a function of frequency. In Fig. 9b we show the difference of the absolute values of the group velocities of waves propagating in opposite directions, $|\Delta v_g| = |\vec{v}_g| - |\overleftarrow{v}_g|$, for each eigenvector, $\mathbf{e}_+$ and $\mathbf{e}_-$. The group velocities associated with the arrow "$\rightarrow$" ("$\leftarrow$") correspond to wave propagation along the $+z$ ($-z$) –direction.

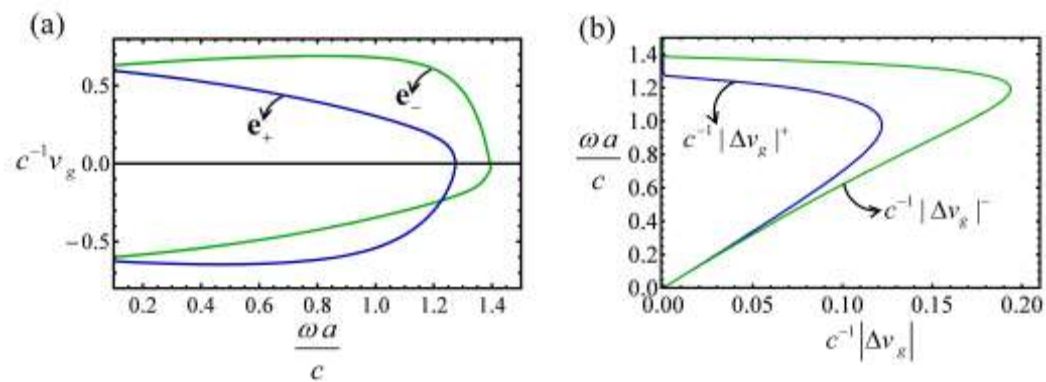

**Fig. 9** (Color online) Group velocities of the Bloch waves. (a) The blue (dark gray) line represents the group velocity of the wave $\mathbf{e}_+$ and the green (light gray) line corresponds to the group velocity of the wave $\mathbf{e}_-$. (b) Difference of the absolute values of the group velocities for counter-propagating waves.

It is interesting to investigate how strong is the spectral asymmetry in a realistic photonic crystal wherein the Tellegen parameter $\kappa$ has magnitude comparable to the magnetoelectric parameter of $Cr_2O_3$ [41]. Hence, next we consider that the constitutive parameters of the Tellegen medium are $\varepsilon = 2$, $\mu = 1$ and $\kappa = 10^{-3}$. The remaining structural and material parameters are the same as in the previous example. The band diagram of this more realistic structure is plotted in the Fig. 10a and the frequency range near the band gap is zoomed in Fig. 10b and Fig. 10c. As seen, the asymmetry is tiny but detectable near the band gaps.

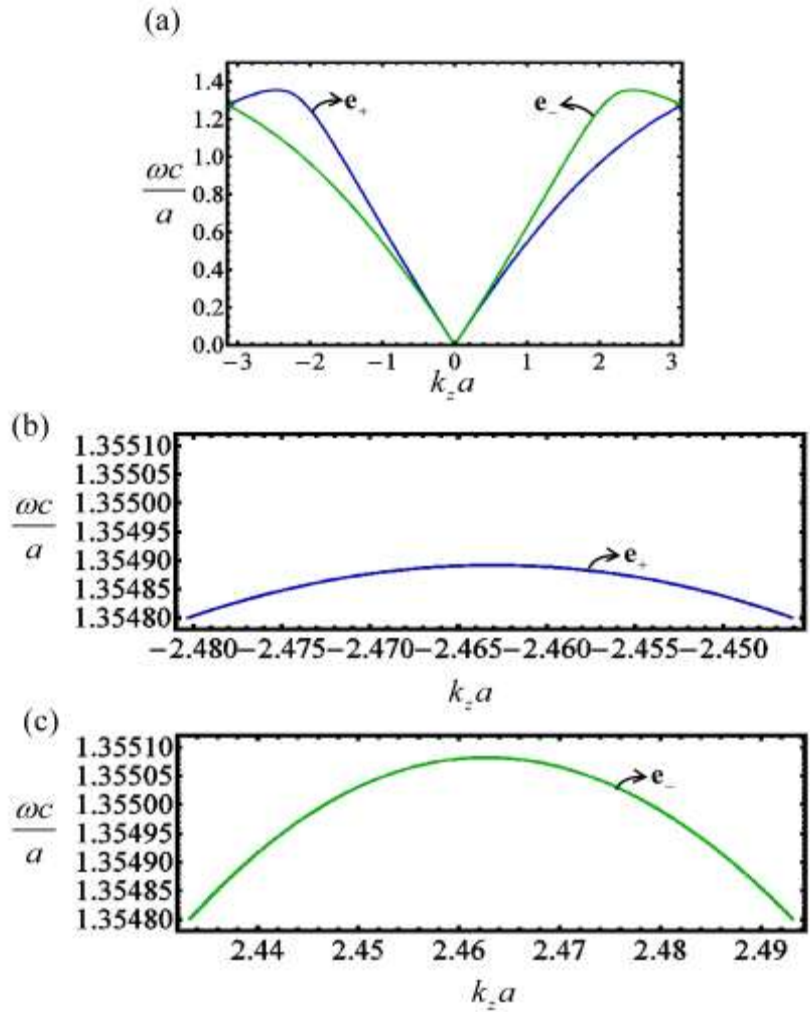

**Fig. 10** (Color online) (a) Band diagram of BI periodic structures formed by a Tellegen material with $\kappa = 10^{-3}$. (b) and (c) Zooms of the band diagram near the band-gap.

## E. Band diagrams for periodic structures with ferrites and chiral media

Finally, we analyze periodic structures formed by nonreciprocal ferrite materials, conventional dielectrics and chiral media. The dispersion of the Bloch waves in ferrite photonic crystals is also found by solving Eq. (22).

In the first example, we suppose that the unit cell is formed by a dielectric slab and a ferrite slab with thicknesses $d_1 = d_2 = 0.5a$. The calculated band diagram is shown in Fig. 11a. The constitutive parameters of the dielectric are $\varepsilon_1 = 2$ and $\mu_1 = 1$ and the dispersion of the lossless ferrite material is modelled by Eq. (17). In this simulation it was assumed that $\omega_R a/c = 4$ and $\omega_m a/c = 2$, $\varepsilon_f = 1$. It is clear from Fig. 11a that for this periodic structure the band diagrams corresponding to the two polarization eigenstates $\mathbf{e}_+$ and $\mathbf{e}_-$ are different. Interestingly, consistent with the discussion at the end of Sect. III, it is seen that a unit cell formed by a single ferrite slab and a dielectric is equivalent to a three material Tellegen unit cell wherein the media are in different equivalence classes [see Sect. IV B]. In particular, it is seen that the band diagrams are symmetric. This property can be readily explained with the help of the transformation theory outlined in Fig. 8. From Fig. 8c it is readily checked that this photonic crystal is invariant under a spatial inversion ($\mathcal{I}_{x,y,z}$) because the dielectric has vanishing chiral and Tellegen parameters. This implies that $\omega(k_z) = \omega(-k_z)$.

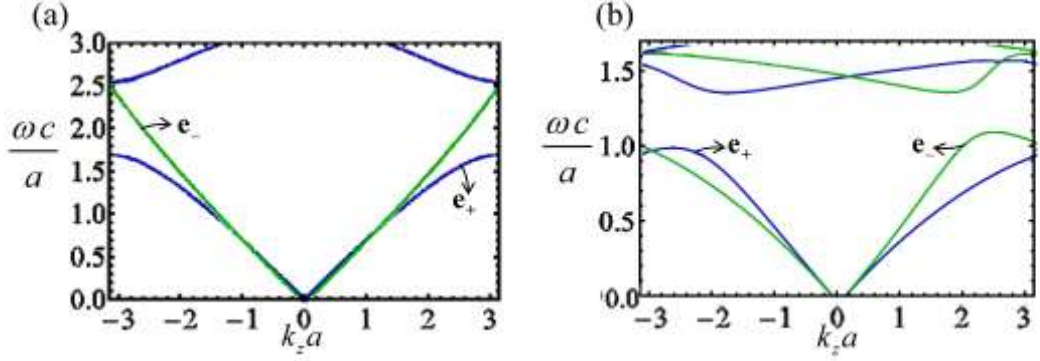

**Fig. 11** (Color online) Band diagram of a stratified periodic structure. (a) Unit cell formed by a dielectric slab and a ferrite slab. (b) Unit cell formed by a chiral medium slab and a ferrite slab.

Interestingly, a photonic crystal whose unit cell is formed by a ferrite slab and a chiral slab is not protected by any of the symmetry transformations considered in Fig. 8. Hence, this suggests that such a configuration can allow for asymmetric band diagrams. This is indeed the case, as shown in Fig. 11b. In this second example, the unit cell contains a chiral slab and a ferrite slab with thicknesses $d_1 = d_2 = 0.5a$. The constitutive parameters of the ferrite material are as in the previous example, while the chiral medium parameters are as in Sect. IV.D. This example and the results of Sect. IV.D further confirm that the role of a single ferrite slab can be mimicked by two Tellegen slabs (one of the Tellegen slabs may be a regular dielectric).

## V. Conclusions

We investigated the scattering by stratified structures formed by Tellegen (e.g. topological insulators or $Cr_2O_3$) or ferrite materials. It was found that if the Tellegen structures cannot be simultaneously reduced under a duality transformation to conventional dielectrics, then it is possible to have asymmetric transmissions of waves incident on opposite sides of the stratified structures. Otherwise, nonreciprocal effects are only manifested in the reflected waves.

The propagation of electromagnetic waves in periodic stratified structures formed by chiral media and Tellegen or ferrite materials was studied. In case of Tellegen photonic crystals, it was proven that in general the eigenmodes are not degenerate for propagation along the stratification direction, except if the structure exhibits a two-fold rotation symmetry (about the *y*-axis) or, alternatively, if all the involved media are in the same equivalence class. Surprisingly, we discovered that notwithstanding the nonreciprocal response of Tellegen media, the photonic band diagrams are always spectrally symmetric. We explained this result by showing that the spectral symmetry is protected by a symmetry transformation that corresponds to the time-reversal operation followed by the *y*-inversion operation.

With the motivation of finding a bi-isotropic crystal wherein the spectral symmetry is broken, we considered periodic stratified structures with both Tellegen and chiral media. The band structure of such crystals can be highly asymmetric and in particular the group velocities of counter-propagating waves are different. We investigated how pronounced this effect can be in photonic crystals formed by a Tellegen material with parameters consistent with those of $Cr_2O_3$ and found that although very tiny the asymmetry is revealed. Furthermore, it was proven that periodic structures formed by chiral media and ferrites also have asymmetric band diagrams. In general, the role of a ferrite slab biased with a static field oriented along the *z*-direction can be mimicked by two juxtaposed Tellegen media slabs with a spontaneous nonreciprocal response.

Because the electromagnetic response of electronic topological insulators may be equivalent to the response of Tellegen media, our findings suggest exciting applications of these materials in novel photonic couplers with asymmetric transmissions. This topic will be further explored in future work.

# Appendix A: Axion electrodynamics and the Tellegen medium

The Maxwell equations in presence of a time independent axion-coupling term ($\vartheta$) are [25]:

$$\nabla \times \mathbf{E} = -\frac{\partial \mathbf{B}}{\partial t}, \tag{A1}$$

$$\nabla \times \frac{\mathbf{B}}{\mu} = \varepsilon \frac{\partial \mathbf{E}}{\partial t} + \mathbf{j}^{(e)} - \nu(\nabla \vartheta \times \mathbf{E}), \tag{A2}$$

where $\nu$ is a coupling constant. Substituting $\nabla \vartheta \times \mathbf{E} = \nabla \times (\vartheta \mathbf{E}) - \vartheta \nabla \times \mathbf{E}$ into Eq. (A.2) and using Eq. (A.1) one obtains:

$$\nabla \times \frac{\mathbf{B}}{\mu} = \varepsilon \frac{\partial \mathbf{E}}{\partial t} + \mathbf{j}^{(e)} - \nu\left(\nabla \times \vartheta \mathbf{E} + \vartheta \frac{\partial \mathbf{B}}{\partial t}\right). \tag{A3}$$

After some manipulations Eq. (A.3) can be rewritten as:

$$\nabla \times \left(\frac{\mathbf{B}}{\mu} + \nu \vartheta \mathbf{E}\right) = \frac{\partial}{\partial t}(\varepsilon \mathbf{E} - \nu \vartheta \mathbf{B}) + \mathbf{j}^{(e)}. \tag{A4}$$

Hence, introducing the electric displacement field vector, $\mathbf{D}$, and the magnetic field vector, $\mathbf{H}$, defined as

$$\mathbf{H} = \frac{\mathbf{B}}{\mu} + \nu \vartheta \mathbf{E}, \quad \mathbf{D} = \varepsilon \mathbf{E} - \nu \vartheta \mathbf{B}, \tag{A5}$$

we get

$$\nabla \times \mathbf{H} = \frac{\partial \mathbf{D}}{\partial t} + \mathbf{j}^{(e)}. \tag{A6}$$

Equation (A.5) is coincident with Eq. (1) of [28]. Clearly, the constitutive relations in Eq. (A.5) can be rewritten as:

$$\mathbf{B} = \mu_0 \mu' \mathbf{H} + c^{-1} \kappa' \mathbf{E}, \quad \mathbf{D} = \varepsilon_0 \varepsilon' \mathbf{E} + c^{-1} \kappa' \mathbf{H}, \tag{A7}$$

where $c^{-1}\kappa' = -\nu\vartheta\mu$, $\varepsilon_0\varepsilon' = \varepsilon + \mu(\nu\vartheta)^2$, $\mu_0\mu' = \mu$ and the refractive index is $n' = \sqrt{\varepsilon'\mu' - \kappa'^2} = \sqrt{\varepsilon\mu/(\varepsilon_0\mu_0)}$. Hence, the axion electrodynamics leads to the constitutive relations of a Tellegen medium. From this result it follows that if the axion parameter $\vartheta$ is piecewise constant in space, then at an interface (i.e. at the surface wherein $\vartheta$ is discontinuous) the tangential electric field (**E**) and the tangential magnetic field (**H**) are required to be continuous [24]. On the other hand, the tangential component of $\dfrac{\mathbf{B}}{\mu}$ is in general discontinuous.

**Appendix B: Transmission matrices for stratified Tellegen media**

In this Appendix we prove that, independent of the number of layers, the transmission matrices associated with Tellegen stratified structures satisfy $\bar{\mathbf{T}}^L = \bar{\mathbf{T}}^R$. To begin with, we consider the simple case wherein the stratified structure is formed by a single Tellegen slab (air-Tellegen-air) with thickness $d_1$. The transmitted electric field vector can be formally calculated based on a (vector) reflection diagram that accounts for the multiple wave reflections/transmissions at the two interfaces. The reflection diagrams for incidence from the left-hand side (propagation L-R) is shown in Fig. 12a. A similar reflection diagram can be constructed for incidence from the right-hand side (propagation R-L) (not shown).

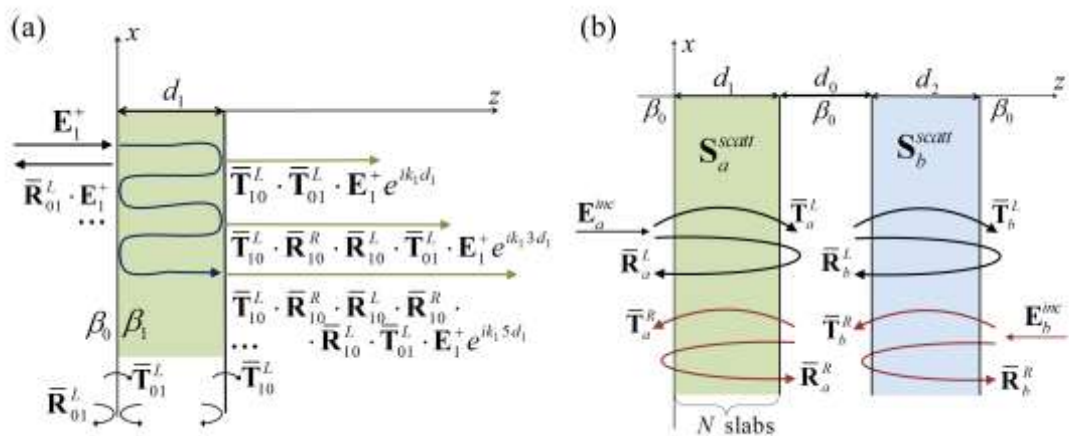

**Fig. 12** (Color online) (a) Reflection diagram for the scattering of a plane wave by a Tellegen slab for wave propagation L-R. (b) A set of $N$-Tellegen slabs is juxtaposed to a single material Tellegen slab. Each structure is characterized by the respective global reflection and transmission matrices.

In both cases, the transmitted electric field vectors $\mathbf{E}_T^L$ and $\mathbf{E}_T^R$ are a superposition of multiple transmitted waves, as shown in the reflection diagram of Fig. 12a. Hence, the total transmitted field for an incoming wave propagating in the L-R direction, $\mathbf{\bar{E}}_T^L$, is given by the following geometric series

$$\begin{aligned}\mathbf{E}_T^L &= \mathbf{\bar{T}}_{10}^L \cdot \mathbf{\bar{T}}_{01}^L \cdot \mathbf{E}_1^+ e^{ik_1 d_1} + \mathbf{\bar{T}}_{10}^L \cdot \left(\mathbf{\bar{R}}_{10}^R \cdot \mathbf{\bar{R}}_{10}^L\right) \cdot \mathbf{\bar{T}}_{01}^L \cdot \mathbf{E}_1^+ e^{ik_1 3 d_1} \\ &+ \mathbf{\bar{T}}_{10}^L \cdot \left(\mathbf{\bar{R}}_{10}^R \cdot \mathbf{\bar{R}}_{10}^L \cdot \mathbf{\bar{R}}_{10}^R \cdot \mathbf{\bar{R}}_{10}^L\right) \cdot \mathbf{\bar{T}}_{01}^L \cdot \mathbf{E}_1^+ e^{ik_1 5 d_1} + ... \\ &= \mathbf{\bar{T}}_{10}^L \cdot \frac{1}{1 - \mathbf{\bar{R}}_{10}^R \cdot \mathbf{\bar{R}}_{10}^L e^{ik_1 2 d_1}} \cdot \mathbf{\bar{T}}_{01}^L \cdot \mathbf{E}_1^+ e^{ik_1 d_1}\end{aligned} \quad (B1)$$

where $\left(\mathbf{\bar{T}}_{01}^L, \mathbf{\bar{T}}_{10}^L\right)$ are the transmission matrices associated with the air-Tellegen and Tellegen-air interfaces, respectively, for a wave propagating in the L-R direction. On the other hand, $\left(\mathbf{\bar{R}}_{10}^L, \mathbf{\bar{R}}_{10}^R\right)$ are the reflection matrices associated with the Tellegen-air interface for a wave propagating in the L-R and R-L directions, respectively. In a similar manner, it is possible to prove that the total transmitted field for an incoming wave propagating in the R-L direction is

$$\begin{aligned}\mathbf{E}_T^R &= \mathbf{\bar{T}}_{10}^R \cdot \mathbf{\bar{T}}_{01}^R \cdot \mathbf{E}_1^- e^{ik_1 d_1} + \mathbf{\bar{T}}_{10}^R \cdot \left(\mathbf{\bar{R}}_{10}^L \cdot \mathbf{\bar{R}}_{10}^R\right) \cdot \mathbf{\bar{T}}_{01}^R \cdot \mathbf{E}_1^- e^{ik_1 3 d_1} \\ &+ \mathbf{\bar{T}}_{10}^R \cdot \left(\mathbf{\bar{R}}_{10}^L \cdot \mathbf{\bar{R}}_{10}^R \cdot \mathbf{\bar{R}}_{10}^L \cdot \mathbf{\bar{R}}_{10}^R\right) \cdot \mathbf{\bar{T}}_{01}^R \cdot \mathbf{E}_1^- e^{ik_1 5 d_1} + ... \\ &= \mathbf{\bar{T}}_{10}^R \cdot \frac{1}{1 - \mathbf{\bar{R}}_{10}^L \cdot \mathbf{\bar{R}}_{10}^R e^{ik_1 2 d_1}} \cdot \mathbf{\bar{T}}_{01}^R \cdot \mathbf{E}_1^- e^{ik_1 d_1}\end{aligned} \quad (B2)$$

where $\left(\mathbf{\bar{T}}_{01}^R, \mathbf{\bar{T}}_{10}^R\right)$ are the transmission matrices for the air-Tellegen and Tellegen-air interfaces, respectively, for a wave propagating in the R-L direction. Thus, from Eqs. (B1) and (B2) the global transmission matrices $\left(\mathbf{\bar{T}}^L, \mathbf{\bar{T}}^R\right)$ for the Tellegen slab satisfy

$$\mathbf{\bar{T}}^L = \mathbf{\bar{T}}_{10}^L \cdot \frac{1}{1 - \mathbf{\bar{R}}_{10}^R \cdot \mathbf{\bar{R}}_{10}^L e^{ik_1 2 d_1}} \cdot \mathbf{\bar{T}}_{01}^L e^{ik_1 d_1}, \quad \mathbf{\bar{T}}^R = \mathbf{\bar{T}}_{10}^R \cdot \frac{1}{1 - \mathbf{\bar{R}}_{10}^L \cdot \mathbf{\bar{R}}_{10}^R e^{ik_1 2 d_1}} \cdot \mathbf{\bar{T}}_{01}^R e^{ik_1 d_1}. \quad (B3)$$

Using arguments analogous to those of Sect. II.B it readily follows that $\mathbf{\bar{T}}_{01}^L, \mathbf{\bar{T}}_{10}^L, \mathbf{\bar{T}}_{01}^R, \mathbf{\bar{T}}_{10}^R, \mathbf{\bar{R}}_{10}^L, \mathbf{\bar{R}}_{10}^R$ are necessarily matrices of the form presented in Eq. (10).

Thus, all the relevant matrices commute and hence to prove that $\bar{\mathbf{T}}^L = \bar{\mathbf{T}}^R$ it suffices to show that $\bar{\mathbf{T}}_{10}^L \cdot \bar{\mathbf{T}}_{01}^L = \bar{\mathbf{T}}_{10}^R \cdot \bar{\mathbf{T}}_{01}^R$. The transmission matrices $\left( \bar{\mathbf{T}}_{01}^L, \bar{\mathbf{T}}_{10}^L, \bar{\mathbf{T}}_{01}^R, \bar{\mathbf{T}}_{10}^R \right)$ may be explicitly written as a function of the characteristic admittance dyadics $\left( \bar{\mathbf{Y}}_0^+, \bar{\mathbf{Y}}_0^-, \bar{\mathbf{Y}}_1^+, \bar{\mathbf{Y}}_1^- \right)$ (defined as explained below Eq. (3)) as

$$\bar{\mathbf{T}}_{01}^L = \left[ \bar{\mathbf{Y}}_0^- + \bar{\mathbf{Y}}_1^+ \right]^{-1} \cdot \left( \bar{\mathbf{Y}}_0^+ + \bar{\mathbf{Y}}_0^- \right), \qquad \bar{\mathbf{T}}_{10}^L = \left[ \bar{\mathbf{Y}}_1^- + \bar{\mathbf{Y}}_0^+ \right]^{-1} \cdot \left( \bar{\mathbf{Y}}_1^+ + \bar{\mathbf{Y}}_1^- \right)$$
$$\bar{\mathbf{T}}_{01}^R = \left[ \bar{\mathbf{Y}}_0^+ + \bar{\mathbf{Y}}_1^- \right]^{-1} \cdot \left( \bar{\mathbf{Y}}_0^- + \bar{\mathbf{Y}}_0^+ \right), \qquad \bar{\mathbf{T}}_{10}^R = \left[ \bar{\mathbf{Y}}_1^+ + \bar{\mathbf{Y}}_0^- \right]^{-1} \cdot \left( \bar{\mathbf{Y}}_1^- + \bar{\mathbf{Y}}_1^+ \right)$$
(B4)

The characteristic admittances are also of the generic form of Eq. (10) and hence because the sums, inverses, and products of such matrices commute, it follows that $\bar{\mathbf{T}}_{10}^L \cdot \bar{\mathbf{T}}_{01}^L = \bar{\mathbf{T}}_{10}^R \cdot \bar{\mathbf{T}}_{01}^R$. This concludes the proof for the case of a single material slab.

Next, we generalize the result $\bar{\mathbf{T}}^L = \bar{\mathbf{T}}^R$ for the case of $N$- juxtaposed Tellegen slabs. The proof is done by induction in the number of slabs. Consider a generic structure formed by $N+1$ slabs, as depicted in Fig. 12b. This structure can be regarded as a collection of $N$ slabs with total thickness $d_1$ and another slab with thickness $d_2$. It is convenient to imagine that the two structures stand in air and are separated by a distance $d_0$. In the end, we will consider the limit of a vanishing $d_0$. The global transmission and reflection matrices for the first structure (i.e., the collection of $N$ slabs) are denoted by $\left( \bar{\mathbf{T}}_a^L, \bar{\mathbf{T}}_a^R, \bar{\mathbf{R}}_a^L, \bar{\mathbf{R}}_a^R \right)$ whereas for the second slab they are $\left( \bar{\mathbf{T}}_b^L, \bar{\mathbf{T}}_b^R, \bar{\mathbf{R}}_b^L, \bar{\mathbf{R}}_b^R \right)$. Because of the induction hypothesis, we can assume that $\bar{\mathbf{T}}_a^L = \bar{\mathbf{T}}_a^R$ and $\bar{\mathbf{T}}_b^L = \bar{\mathbf{T}}_b^R$. Next, we compute the global transmission matrices for the set of $N+1$-slabs.

To do this, we use again reflection diagrams analogous to those of Fig. 12. In this manner, we find that for an incoming wave propagating in the L-R direction the transmitted electric field vector $\bar{\mathbf{E}}_T^{\prime L}$ is given by

$$\mathbf{E}_T'^L = \bar{\mathbf{T}}_b^L \cdot \frac{1}{1 - \bar{\mathbf{R}}_a^R \cdot \bar{\mathbf{R}}_b^L e^{ik_0 2d_0}} \cdot \bar{\mathbf{T}}_a^L \cdot \mathbf{E}_a^{inc} e^{ik_0 d_0} \tag{B5}$$

whereas for an incoming wave propagating in the R-L direction one obtains

$$\mathbf{E}_T'^R = \bar{\mathbf{T}}_a^R \cdot \frac{1}{1 - \bar{\mathbf{R}}_b^L \cdot \bar{\mathbf{R}}_a^R e^{ik_0 2d_0}} \cdot \bar{\mathbf{T}}_b^R \cdot \mathbf{E}_b^{inc} e^{ik_0 d_0} . \tag{B6}$$

Thus, letting the thickness between the structures to approach zero, $d_0 \to 0$, using the induction hypothesis, and the fact that because of the isotropy of the involved materials all the matrices are necessarily of the form presented in Eq. (10) and thus commute, it follows that the global transmission coefficients for the set of the $N+1$ juxtaposed slabs $\left(\bar{\mathbf{T}}_c^L, \bar{\mathbf{T}}_c^R\right)$ satisfy:

$$\bar{\mathbf{T}}_c^L = \bar{\mathbf{T}}_b^L \cdot \frac{1}{1 - \bar{\mathbf{R}}_a^R \cdot \bar{\mathbf{R}}_b^L} \cdot \bar{\mathbf{T}}_a^L = \bar{\mathbf{T}}_a^R \cdot \frac{1}{1 - \bar{\mathbf{R}}_b^L \cdot \bar{\mathbf{R}}_a^R} \cdot \bar{\mathbf{T}}_b^R = \bar{\mathbf{T}}_c^R . \tag{B7}$$

By definition, the global transmission matrices are such that $\mathbf{E}_T'^L = \bar{\mathbf{T}}_c^L \cdot \mathbf{E}_a^{inc}$ and $\mathbf{E}_T'^R = \bar{\mathbf{T}}_c^R \cdot \mathbf{E}_b^{inc}$. Thus, this result proves that the induction hypothesis also holds for the set of $N+1$-slabs, and thus it is valid for arbitrary $N$, as we wanted to show.

## Appendix C: Duality transformations

In this Appendix we briefly review duality transformations applied to Tellegen media. Duality transformations are linear mappings of the electromagnetic fields of the form [42]:

$$\begin{pmatrix} \mathbf{E}_d \\ \eta_0 \mathbf{H}_d \end{pmatrix} = \bar{\mathbf{S}} \cdot \begin{pmatrix} \mathbf{E} \\ \eta_0 \mathbf{H} \end{pmatrix}, \text{ with } \bar{\mathbf{S}} = \begin{pmatrix} s_{11} & s_{12} \\ s_{21} & s_{22} \end{pmatrix}, \tag{C1}$$

where $\bar{\mathbf{S}}$ is a 2×2 real-valued matrix with constant elements (independent of the spatial coordinates) and $s_{11}$, $s_{12}$, $s_{21}$ and $s_{22}$ are real-valued parameters. It is well-known that the duality transformed fields $\mathbf{E}_d$ and $\mathbf{H}_d$ are solutions of the Maxwell's

equations in a transformed structure described by the transformed material matrix [39]:

$$\bar{\mathbf{M}}_d = \det(\bar{\mathbf{S}})(\bar{\mathbf{S}}^{-1})^{\mathrm{T}} \cdot \bar{\mathbf{M}} \cdot \bar{\mathbf{S}}^{-1}, \qquad (C2)$$

where $\bar{\mathbf{M}} = \begin{pmatrix} \varepsilon & \kappa \\ \kappa & \mu \end{pmatrix}$ is the material matrix of a Tellegen medium defined so that

$$\begin{pmatrix} \mathbf{D}/\varepsilon_0 \\ \mathbf{B}c \end{pmatrix} = \bar{\mathbf{M}} \cdot \begin{pmatrix} \mathbf{E} \\ \eta_0 \mathbf{H} \end{pmatrix}. \qquad (C3)$$

In a previous work, we demonstrated that electromagnetic structures formed by Tellegen media can be transformed, in some conditions, into simpler structures formed by simple isotropic media using a duality mapping [39]. This can be rather useful because if a solution for the duality transformed problem can be found, then the solution of the original problem can be readily obtained by applying an inverse duality mapping [42]-[44]. For example, if a Tellegen photonic crystal is reducible to a crystal formed by only simple isotropic media using a duality transformation, then the dispersion diagrams of the original structure are exactly the same as the diagrams of the transformed photonic crystal [39]. It was shown in [39] that typically three different Tellegen media cannot be transformed into simple isotropic media using the same duality transformation, or in other words, they do not belong to the same Tellegen equivalence class [39]. These results were developed using a geometric approach wherein the Tellegen equivalence classes reducible to the simple isotropic media class are represented by circles in the Riemann sphere [39]. For more details, a reader is referred to our previous work [39].